\documentclass[aps,prb,twocolumn,superscriptaddress]{revtex4-1}

\usepackage{amsfonts}
\usepackage{amssymb}
\usepackage{amsmath}
\usepackage{upgreek}
\usepackage{color} 
\usepackage{lineno}
\usepackage[percent]{overpic}
\usepackage[export]{adjustbox}
\usepackage{stackengine}
\usepackage{appendix}

\usepackage{graphicx}
\usepackage{hyperref}
\hypersetup{colorlinks=true,
            citecolor=[rgb]{0,0,0.502},
            urlcolor=[rgb]{0,0,0.502},
            linkcolor=[rgb]{0,0,0.502}}

\newcommand*\patchAmsMathEnvironmentForLineno[1]{%
  \expandafter\let\csname old#1\expandafter\endcsname\csname #1\endcsname
  \expandafter\let\csname oldend#1\expandafter\endcsname\csname end#1\endcsname
  \renewenvironment{#1}%
     {\linenomath\csname old#1\endcsname}%
     {\csname oldend#1\endcsname\endlinenomath}}%
\newcommand*\patchBothAmsMathEnvironmentsForLineno[1]{%
  \patchAmsMathEnvironmentForLineno{#1}%
  \patchAmsMathEnvironmentForLineno{#1*}}%
\AtBeginDocument{%
\patchBothAmsMathEnvironmentsForLineno{equation}%
\patchBothAmsMathEnvironmentsForLineno{align}%
\patchBothAmsMathEnvironmentsForLineno{flalign}%
\patchBothAmsMathEnvironmentsForLineno{alignat}%
\patchBothAmsMathEnvironmentsForLineno{gather}%
\patchBothAmsMathEnvironmentsForLineno{multline}%
}

\newcommand{\mytitle}{Self-similar dynamics of order parameter fluctuations in pump-probe experiments}

\newcommand{\MIT}{Massachusetts Institute of Technology, Department of Physics, Cambridge, Massachusetts 02139, USA.}
\newcommand{\Harvard}{Department of Physics, Harvard University, Cambridge, Massachusetts 02138, USA.}

\begin{document}

\title{\mytitle}

\author{Pavel~E.~Dolgirev}
\email[Correspondence to: ]{p\_dolgirev@g.harvard.edu}
\author{Marios~H.~Michael}
\affiliation{\Harvard}
\author{Alfred~Zong}
\affiliation{\MIT}
\author{Nuh~Gedik}
\affiliation{\MIT}
\author{Eugene~Demler}
\affiliation{\Harvard}

\def\mean#1{\left< #1 \right>}

\date{\today}

\begin{abstract}
Upon excitation by a laser pulse, broken-symmetry phases of a wide variety of solids demonstrate similar order parameter dynamics characterized by a dramatic slowing down of relaxation for stronger pump fluences. Motivated by this recurrent phenomenology, we develop a simple non-perturbative effective model for photoinduced dynamics of collective bosonic excitations. We find that as the system recovers after photoexcitation, it shows universal prethermalized dynamics manifesting a power-law, as opposed to exponential, relaxation, explaining the slowing down of the recovery process. For strong quenches, long-wavelength over-populated transverse modes dominate the long-time dynamics; their distribution function exhibits universal scaling in time and space, whose universal exponents can be computed analytically. Our model offers a unifying description of order parameter fluctuations in a regime far from equilibrium, and our predictions can be tested with available time-resolved techniques.

\end{abstract}

\maketitle


\section{Introduction}
\label{sec:intro}

The concept of universality plays a central role in the theory of equilibrium phase transitions because it allows us to reduce a plethora of experimentally studied systems to a few fundamental classes~\cite{Goldenfeld1992}. For systems far from equilibrium, the notion of universality~\cite{hohenberg1977theory} is relatively unexplored and has recently emerged as an active field~\cite{schwarz1988three,micha2003relativistic,micha2004turbulent,berges2008nonthermal,berges2011strong,nowak2012nonthermal,orioli2015universal,bhattacharyya2019universal,nowak2011superfluid}, partially motivated by recent progress in ultracold-atom~\cite{erne2018universal,prufer2018observation,eigen2018universal,navon2016emergence} and ultrafast pump-probe experiments~\cite{mitrano2019}. In a non-equilibrium context, one of the dramatic manifestations of universality is the emergence of the self-similar evolution of correlation functions~\cite{barenblatt1996scaling,zakharov2012kolmogorov}. In particular, after a strong perturbation, the transient equal-time two-point correlation function $D(|{\bf x} - {\bf y}|, t)$ might depend only on a single evolving length scale $\xi(t)$ and two universal functions: 
\begin{equation}
    D(|{\bf x} - {\bf y}|, t) = g(t) f(|{\bf x} - {\bf y}|/\xi(t)). \label{eqn:first}
\end{equation}
Functional forms of $f(x)$ and $g(t)$ depend neither on microscopic parameters nor on initial conditions. Typical equations of motion, often a complex system of partial integro-differential equations, represent an interplay between many degrees of freedom such as quasiparticles, order parameter (OP), phonons and/or magnons. If these equations allow for the above self-similar form, the analysis might reduce to just a few differential equations, which is particularly appealing since it eases the interpretation of the involved evolution. From a physical standpoint, the self-similarity suggests that there exists a stabilization-like mechanism responsible for this form. 

\begin{figure}
    \centering
    \includegraphics[width=1\linewidth]{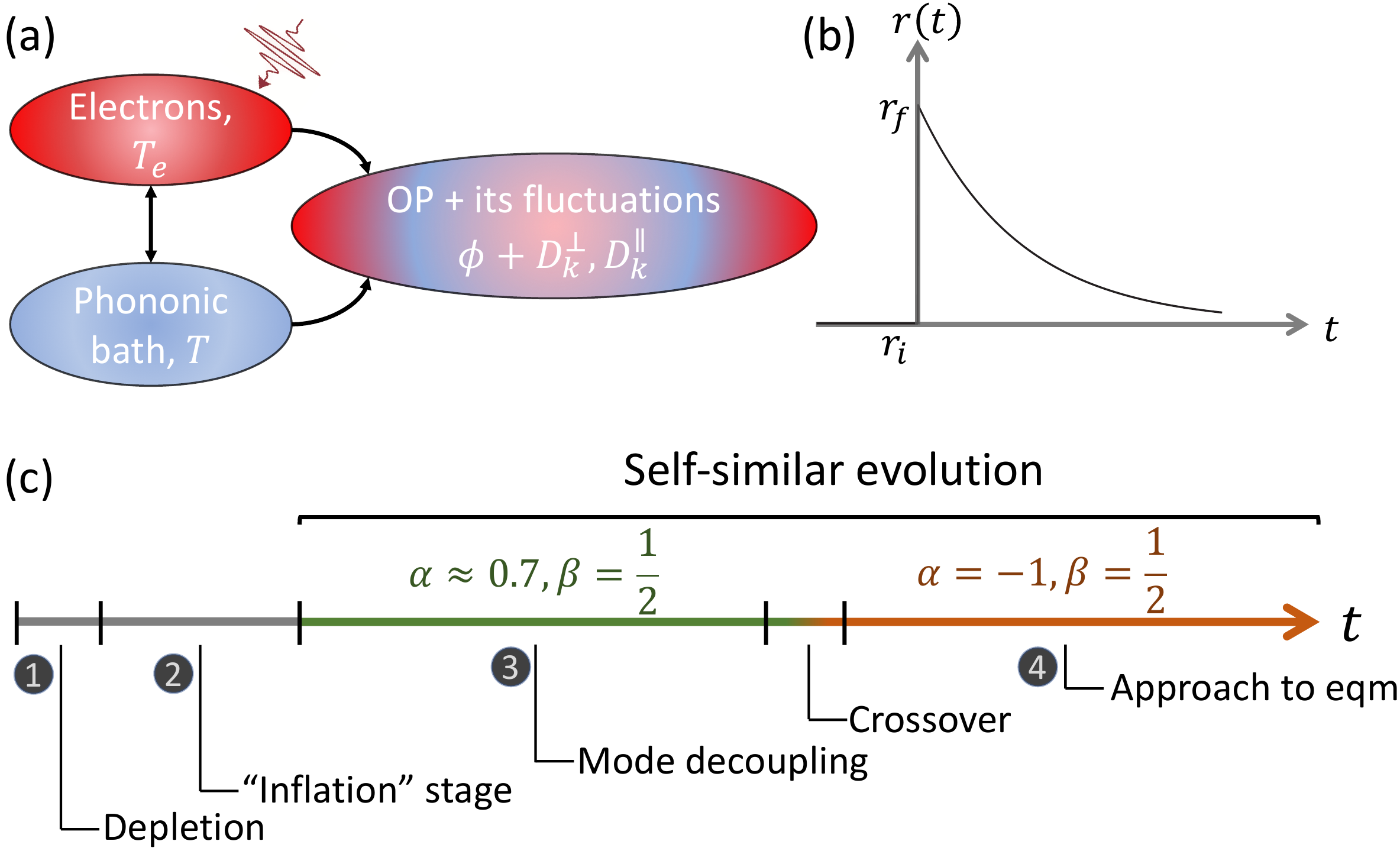}
    \caption{(a)~Schematics of a non-equilibrium state: electrons (red) and the phononic bath (blue) are thermal with temperatures $T_e(t)$ and $T$, respectively; the OP subsystem (mixed colors) is not assumed to be thermal. (b)~Time evolution of the Landau coefficient $r(t)$ [Eq.~\eqref{eqn::quench}]. It mimics a photoexcitation event in (a). (c)~Schematics of dynamical stages. During stages 3 and 4, the system exhibits self-similarity. Green and orange distinguish the scaling exponents, $\alpha$ and $\beta$ [Eq.~\eqref{eqn::exponents}], in these two stages.}
    \label{Fig1}
\end{figure}

Recent pump-probe experiments observed several recurrent phenomena that hint at the existence of universality in the out-of-equilibrium context. In particular, common features have been observed in the dynamics of the OP and low-energy collective excitations in charge-density-wave (CDW) compounds~\cite{tomeljak2009dynamics,schmitt2011ultrafast,zong2019evidence,kogar2019light,zong2019dynamical,ruan2019,MohrVorobeva2011,laulhe2017ultrafast,vogelgesang2018phase,Huber2014,mitrano2019,storeck2019hot}, putative excitonic insulators~\cite{hellmann2012time,okazaki2018photo}, magnetically-ordered systems~\cite{Kirilyuk2010,dean2016ultrafast}, and systems that exhibit several intertwined orders \cite{mariager2014structural,Beaud2014}. Common phenomenology in these materials includes: (i)~The recovery of a photo-suppressed OP takes longer at stronger pump pulse fluence; (ii)~The amplitude of the OP restores faster than the phase, exhibiting a separation of timescales; (iii)~Related to (ii), peaks in diffraction experiments remain broadened compared to equilibrium shape long after photoexcitation, showing prolonged suppression of long-range phase coherence. For example, all of these features have been observed in the same set of experiments~\cite{zong2019evidence} on LaTe$_3$, an incommensurate CDW material. Below, for concreteness, we primarily focus our otherwise generic theoretical formalism on this material but expect that our conclusions should qualitatively apply to other systems as well. We note, however, that quantitative results may be different depending on details of specific systems. For example, in the case of ferromagnetic systems, our analysis should be modified to include the conserved character of the OP~\cite{hohenberg1977theory}.

A common approach to describing many-body dynamics in symmetry-broken states is based on the so-called three-temperature model (3TM)~\cite{tao2013anisotropic,mansart2010ultrafast,perfetti2007ultrafast,dolgirev2019amplitude,johnson2017watching}. In this framework, a non-equilibrium state is characterized by assigning different temperatures to different subsystems, such as electrons, phonons, and OP degrees of freedom~\cite{Remark1}. Upon photoexcitation, most incoming light is absorbed by electrons, instantaneously increasing the electronic temperature, $T_e$. The introduction of $T_e(t)$ is justified provided we are only interested in phononic timescales sufficiently exceeding the fast electron-electron scattering time. Subsequent dynamics corresponds to energy exchange between hot electrons and the other two subsystems. In this process, it is often assumed that the lattice heating is negligible because the lattice heat capacity at room temperature is several orders in magnitude larger than that of electrons. Even though the 3TM suggests an intuitive picture about the interplay among different subsystems, it often lacks theoretical justification. A particularly questionable assumption of the 3TM is that one may assign a temperature to the OP degrees of freedom. Indeed, the laser pulse can easily excite low-energy, low-momenta Goldstone modes that interact weakly with each other and with other degrees of freedom. It is, thus, essential to describe the transient state of the OP degrees of freedom more accurately, without assuming thermalization.

In the present paper, we go beyond the 3TM and formulate a general theory of out-of-equilibrium OP correlations to account for potentially non-thermal states of the OP subsystem -- see Fig.~\ref{Fig1}a. Our theory focuses on non-linear dynamics of  collective bosonic excitations. This should be contrasted to earlier work on the relaxation of quasiparticles in superconductors, in which recombination dynamics can lead to faster relaxation rates for higher quasiparticle densities~\cite{gedik2004single,kusar2008controlled,prasankumar2016optical} (see, however, Ref.~\onlinecite{boschini2018collapse}). Within our effective bosonic model, we find that upon photoexcitation, the system passes through four dynamical stages outlined in Fig.~\ref{Fig1}c. For a strong quench, not only is the OP subsystem far from being thermal but overpopulated slow Goldstone modes fully dominate the intrinsic evolution at long times. Even more strikingly, in the last two dynamical stages in Fig.~\ref{Fig1}c, the distribution function of these modes exhibits self-similar evolution as in Eq.~\eqref{eqn:first}. These findings provide an intuitive physical interpretation of the mentioned phenomenology, as we elucidate below.

The paper is organized as follows. In Sec.~\ref{sec:model}, we develop the aforementioned model of pump-probe experiments. Within this framework, the response to an applied laser pulse and the four dynamical stages are elaborated in Sec.~\ref{sec:stages}. Section~\ref{sec:Universality} is devoted to the universality present in the last two dynamical stages. Section~\ref{sec:discussion} contains further discussion on the applicability of the present theory to real experiments.

\section{Theoretical framework}
\label{sec:model}

Spontaneous symmetry breaking (SSB) is described by the time-dependent Landau-Ginzburg formalism (model-A in Ref.~\onlinecite{hohenberg1977theory}):
\begin{eqnarray}
    \frac{d \phi_{\alpha} ({\bf x}, t)}{dt} = -\Gamma\frac{\delta {\cal F}}{\delta \phi_{\alpha}({\bf x}, t)} + \eta_{\alpha}({\bf x}, t). \label{eqn::model_A}
\end{eqnarray}
Here $\phi_\alpha$ is an $N$-component vector of real fields representing the OP. The free energy functional reads
\begin{eqnarray}
    {\cal F}[{\bf \phi}] = \int d^3 {\bf x} \left[ \frac{r}{2} \phi_{\alpha}^2 + \frac{K}{2} (\nabla \phi_{\alpha})^2 + u (\phi_{\alpha}^2)^2   \right],\label{eqn::F}
\end{eqnarray}
and $\eta_\alpha$ represents the noise originating from the phononic bath at temperature $T$:
\begin{eqnarray}
    \mean{\eta_{\alpha}({\bf x}, t)\eta_{\beta}({\bf x'}, t') } = 2T\Gamma \delta_{\alpha,\beta}\delta({\bf x} - {\bf x'}) \delta(t-t'). \label{eqn::noise}
\end{eqnarray}
Here $r,\, K,\, u,$ and $\Gamma$ are the model parameters. For homogeneous quenches, without loss of generality, we assume that SSB occurs along the first direction: $\phi(t) = \mean{\phi_{1}({\bf x}, t)}$. Associated with the OP are longitudinal (Higgs modes) $D^{\parallel}_{{\bf k}}(t)\equiv \mean{\phi_1({\bf k}; t)\phi_1(-{\bf k}; t) }_c$ and transverse (Goldstone modes) $D^{\perp}_{{\bf k}}( t)\equiv \mean{\phi_{\alpha\neq1}({\bf k}; t)\phi_\alpha(-{\bf k}; t) }_c$ correlation functions. The model-A formalism in Eqs.~\eqref{eqn::model_A}--\eqref{eqn::noise} can be conveniently rewritten in terms of the Fokker-Planck equation:
\begin{eqnarray}
     \partial_t {\cal P} = T\Gamma \sum_{{\bf k},\alpha} \frac{\delta}{\delta \phi_{\alpha,{\bf k}}}\left[ \frac{{\cal P}}{T} \frac{\delta {\cal F}}{\delta \phi_{\alpha,-{\bf k}}} + \frac{\delta {\cal P}}{\delta \phi_{\alpha,-{\bf k}}} \right],\label{eqn::F_P}
\end{eqnarray}
where ${\cal P}([\phi], t)$ is the probability distribution functional of space-dependent field configurations $\phi_\alpha({\bf x})$. To the leading order in $1/N$, ${\cal P}([\phi], t)$ is Gaussian, implying that the OP $\phi(t)$ and the correlators $D^\parallel_{\bf k}(t),\,D^\perp_{\bf k}(t)$ form a closed set of dynamical variables. The self-consistent equations of motion read~\cite{mazenko1985instability,zannetti1993transverse} (see Appendix~\ref{appendix:A} for a derivation)
\begin{eqnarray}
\frac{d \phi(t)}{dt} &=& -\Gamma\,   r_{\rm eff}\, \phi, \label{eqn::dt_phi} \\
\frac{d D^{\perp}_{{\bf k}}(t)}{dt} &=& 2T\Gamma - 2\Gamma (K  {\bf k}^2 + r_{\rm eff})D^{\perp}_{{\bf k}},\label{eqn::dt_Dperp}\\
\frac{d D^{\parallel}_{{\bf k}}(t)}{dt} &=& 2T\Gamma - 2\Gamma (K  {\bf k}^2 + r_{\rm eff} + 8u\phi^2)D^{\parallel}_{{\bf k}}\label{eqn::dt_Dpar}.
\end{eqnarray}
Here the self-consistent ``mass''-term is defined as
\begin{equation}
    r_{\rm eff}(t) = r(t) + 4u \left( \phi^2 + n_{\rm tot}^\parallel + (N-1) n_{\rm tot}^\perp \right),\label{eqn::r_eff}
\end{equation}
where $ n_{\rm tot}^{\perp(\parallel)} \equiv \int^{\Lambda} \frac{d^3 {\bf q}}{(2\pi)^3} D^{\perp(\parallel)}_{{\bf q}}$, $\Lambda$ is the UV-cutoff. Note that quantities such as energy or total number of excitations are not conserved due to the external bath.

The bath, cf. Eq.~\eqref{eqn::noise}, will also always result in the thermalization of the system, in contrast to quenches in the isolated $O(N)$ model, where, to the leading in $1/N$ order, the system does not demonstrate equilibration~\cite{chandran2013equilibration,sciolla2013quantum,chiocchetta2015short,maraga2015aging,chiocchetta2016short}. In that case, thermalization occurs only after subleading corrections are taken into consideration~\cite{berges2004introduction}. As such, the presence of the external bath in our case motivates us to disregard these subleading corrections here.

From the equations of motion, we obtain the equilibrium correlators:
\begin{eqnarray}
D^{\parallel}_{\bf k} = \frac{T}{Kk^2+8u\phi^2 + r_{\rm eff}},\, D^{\perp}_{\bf k} = \frac{T}{K {\bf k}^2 + r_{\rm eff}}. \label{eqn::D_eq}
\end{eqnarray}
This result is a manifestation of the equipartition theorem. In the symmetry broken phase, where $r_{\rm eff} = 0$ and $\phi \neq 0$, we observe that the OP equilibrium value $\phi$ is affected by the thermal fluctuations, cf. Eq.~(\ref{eqn::r_eff}). The transverse correlation length $\xi_\perp\propto r_{\rm eff}^{-1/2}$ is divergent. In the disordered phase, $r_{\rm eff} \neq 0$ and $\phi = 0$, the transverse and longitudinal correlations are not distinguishable.

A useful point of view on the above approximations is as follows. The equations of motion~\eqref{eqn::dt_phi}-\eqref{eqn::r_eff} are equivalent to
\begin{eqnarray}
\frac{d\delta \phi^{\perp}_{\bf k} (t)}{dt} &=& -\Gamma (K  {\bf k}^2 + r_{\rm eff})\phi^{\perp}_{\bf k} + \eta^{\perp}_{\bf k}(t),\label{eqn::perp_k}\\
   \frac{d\delta \phi^{\parallel}_{\bf k} (t)}{dt} &=& -\Gamma (K  {\bf k}^2 + r_{\rm eff} + 8u\phi^2)\phi^{\parallel}_{\bf k} + \eta^{\parallel}_{\bf k}(t),\label{eqn::par_k}
\end{eqnarray}
where $\delta \phi_{\bf k}^{\alpha}$ represents the fluctuating part of the corresponding Fourier mode $\phi^{\alpha}_{\bf k}$. We observe that each of the fluctuating modes lives in an effectively parabolic potential, $\mean{\delta \phi_{\bf k}^{\alpha}} = 0$, and the noise term establishes the equilibrium variances given by Eq.~\eqref{eqn::D_eq}.

We now formulate the quenching protocol. For simplicity, we assume that the electronic temperature $T_e$ cools down to the equilibrium value $T$ with a constant rate $\tau_{\rm QP}$ defined by the electron-phonon coupling. This assumption seems to be not too crude for LaTe$_3$, where the Fermi surface is only partially gaped, and the excited by the laser pulse electrons can relax through a gapless channel~\cite{zong2019evidence,dolgirev2019amplitude}. For a different situation, our framework should be straightforwardly extended. In the usual Landau-Ginzburg theory, the coefficient $r(T_e)$ depends linearly on $T_e$. To mimic a photoexcitation event, we therefore impose the following dynamics on $r(t)$ (Fig.~\ref{Fig1}b):
\begin{align}
    r(t) = r_i + \theta(t)  \exp{(-t/\tau_{\rm QP})} \times (r_f-r_i), \label{eqn::quench}
\end{align}
where $\theta(t)$ is the Heaviside theta function, $r_i$ is the pre-pulse value chosen such that $\phi \neq 0$, and $(r_f - r_i)$ characterizes the pulse strength. Below we focus on time delays much beyond $\tau_{\rm QP}$.

In the next section, we apply the above theoretical framework to investigate the intrinsic dynamics after the arrival of a laser pulse. Our conclusions are not specific to the choice of model parameters, but in the numerical simulations below, we fix them to roughly mimic the experimental results in Ref.~\onlinecite{zong2019evidence}. There, the values of $\Gamma$ and $\tau_{\rm QP}$ are expected to be similar.

\section{Evolution after photoexcitation}
\label{sec:stages}

\begin{figure}[t!]
\centering
\includegraphics[width=1\linewidth, center]{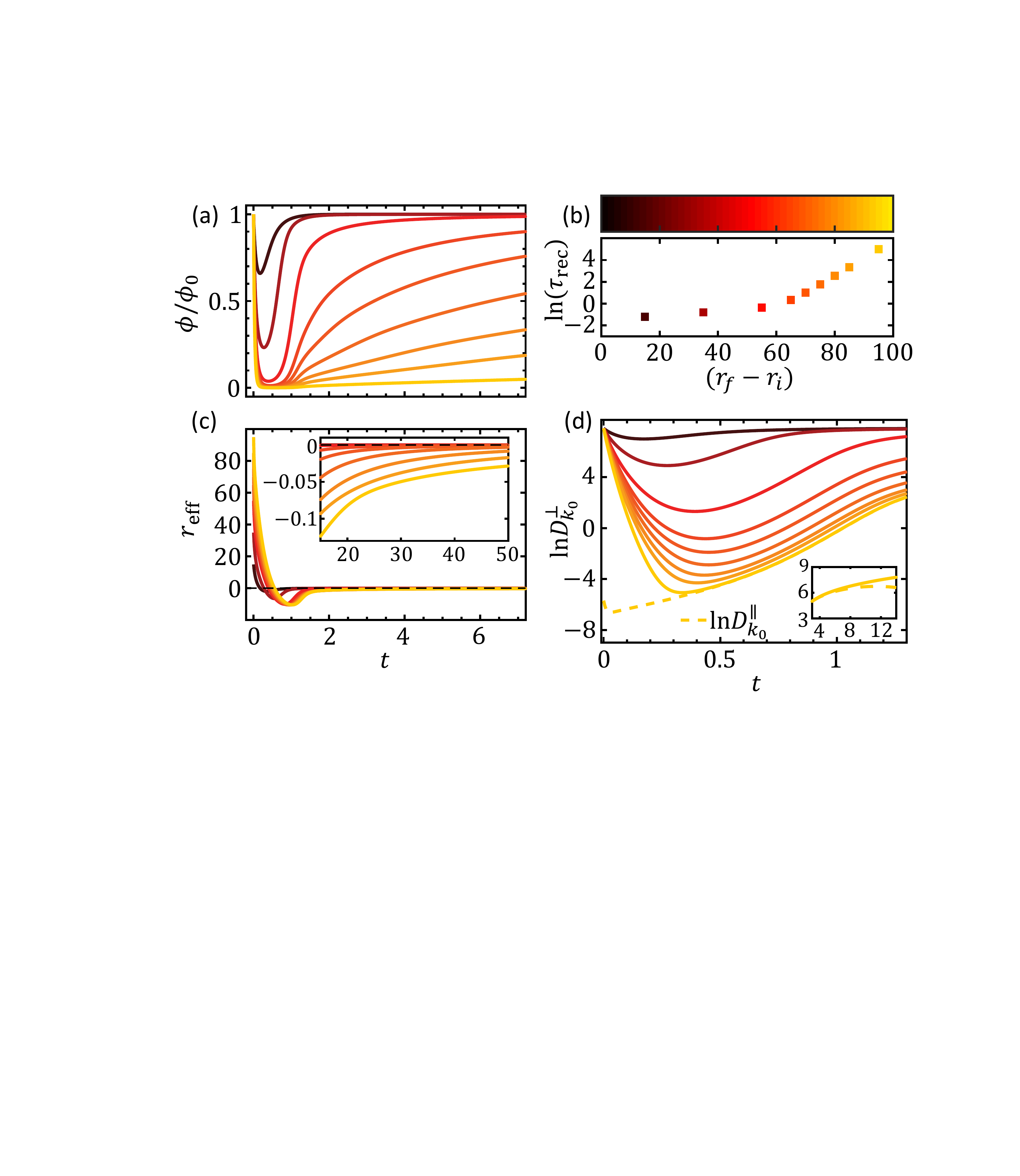}
\caption{Intrinsic dynamics for different quench strengths, $(r_f-r_i)$. (a)~Time dependence of the OP, $\phi(t)$, normalized by its pre-pulse value $\phi_0$. At long times, $(\phi(t) - \phi_0)\sim t^{-d_\phi}$ with $d_\phi = \frac{3}{2}$. (b)~OP recovery time $\tau_{\rm rec}$. (c)~Dynamics of $r_{\rm eff}(t)$. Initially, large positive $r_{\rm eff}$ is suppressed and becomes negative; then it slowly restores 
as $r_{\rm eff} \sim t^{-d_r}$ with $d_r = \frac{5}{2}$
to zero. Inset: Zoomed-in view on the long-time tails. (d)~Evolution of $D^{\perp}_{k_0}$, where $k_0 = \frac{2\pi}{L}$ is the lowest wave vector used in our calculations ($L = 1000$). For a strong pulse, initially $D^{\perp}_{k_0}$ is suppressed to almost zero; after $r_{\rm eff}$ changes sign, it exponentially proliferates. Dotted line corresponds to $D^{\parallel}_{k_0}$ for the strongest pulse considered. $D^{\parallel}_{k_0}$ and $D^{\perp}_{k_0}$ very soon merge into a single curve, indicating that the OP is melted. Inset: longer time dynamics for the strongest pulse; 
$D^{\parallel}_{k_0}$ and $D^{\perp}_{k_0}$ become distinguishable once the OP value $\phi(t)$ becomes appreciable. Throughout the paper, we use the following parameters:  $K = u = 1$, $N = 4$, $\Lambda = \pi$, $\Gamma = 0.5$, $\tau_{\rm QP} = 0.3$, $r_i = -15$, $T = 0.1$. All panels share the same color scale in (b) for the quench strengths.
\label{Fig3}
}
\end{figure}

Upon photoexcitation, cf. Eq.~\eqref{eqn::quench}, the system passes through four dynamical stages (Fig.~\ref{Fig1}c) -- (i) depletion, (ii) inflation, (iii) mode decoupling and (iv) relaxation to the thermal equilibrium. We cover each of them below.

In Fig.~\ref{Fig3}a, we show numerical results for the dynamics of the OP, $\phi(t)$. For a weak pump, $\phi(t)$ becomes slightly suppressed and then quickly recovers to the initial value $\phi_0$. This should be contrasted to the case of a strong pulse, for which initially the OP becomes strongly suppressed and then goes through a long recovery process. The recovery takes longer for stronger pulses (Fig.~\ref{Fig3}b). This slowing-down is due to the power-law dynamics $\delta \phi(t)\equiv (\phi(t) - \phi_0) \sim t^{-d_\phi}$ with $d_\phi = \frac{3}{2}$, which we further discuss in the next section. 

In Fig.~\ref{Fig3}c, we plot the evolution of $r_{\rm eff}(t)$. Upon arrival of a laser pulse, $r_{\rm eff}(+0) = (r_f-r_i)$. This large initial value first decreases due to the time evolution of the ``bare value'' of $r(t)$, cf. Eq.~\eqref{eqn::quench}, and later, at $t \gtrsim \tau_{\rm QP}$, due to the dynamics of the OP and collective modes described by Eqs.~\eqref{eqn::dt_phi}--\eqref{eqn::dt_Dpar}. Even though $r(t)$ returns to its equilibrium value $r_i$ during a relatively short time $\tau_{\rm QP}$, dynamics of $r_{\rm eff}$ occurs over much longer time scale where it even changes sign (Fig.~\ref{Fig3}c). We find that long-time evolution of $r_{\rm eff}\sim t^{-d_r}$ is power-law-like with $d_r = \frac{5}{2}$. For the fluctuating modes $\delta \phi_{\bf k}^{\alpha}$, a large value of $r_{\rm eff}$ implies that each of the effective parabolic potentials becomes initially steeper, and, as such, the noise term in Eq.~\eqref{eqn::noise} tends to depopulate these modes (Fig.~\ref{Fig3}d). Therefore, the first stage -- \emph{depletion} -- is characterized by suppression of the OP and correlations $D^{\perp}_{\bf k}$ and $D^{\parallel}_{\bf k}$.

The second stage -- \emph{inflation} -- starts when $r_{\rm eff}$ changes its sign. A negative $r_{\rm eff}$ implies that each of the effective parabolic potentials becomes shallower or, as the case for the low-momenta transverse modes, can even become inverted. Therefore, during the inflation, population in each of the modes proliferates, most dramatically for the low-momenta modes (Fig.~\ref{Fig3}d).

For a strong quench and at the time when the OP becomes completely suppressed, the longitudinal and transverse correlations are no longer distinguishable (Fig.~\ref{Fig3}d). This parallels the disordered phase in equilibrium situation. As the OP develops, these modes start to separate. We will associate the end of the inflation stage with the time when $D^{\parallel}_{{\bf k} = 0}$ reaches its maximum value (see inset in Fig.~\ref{Fig3}d, dashed curve).

\begin{figure}[t!]
    \centering
    \includegraphics[width=1\linewidth]{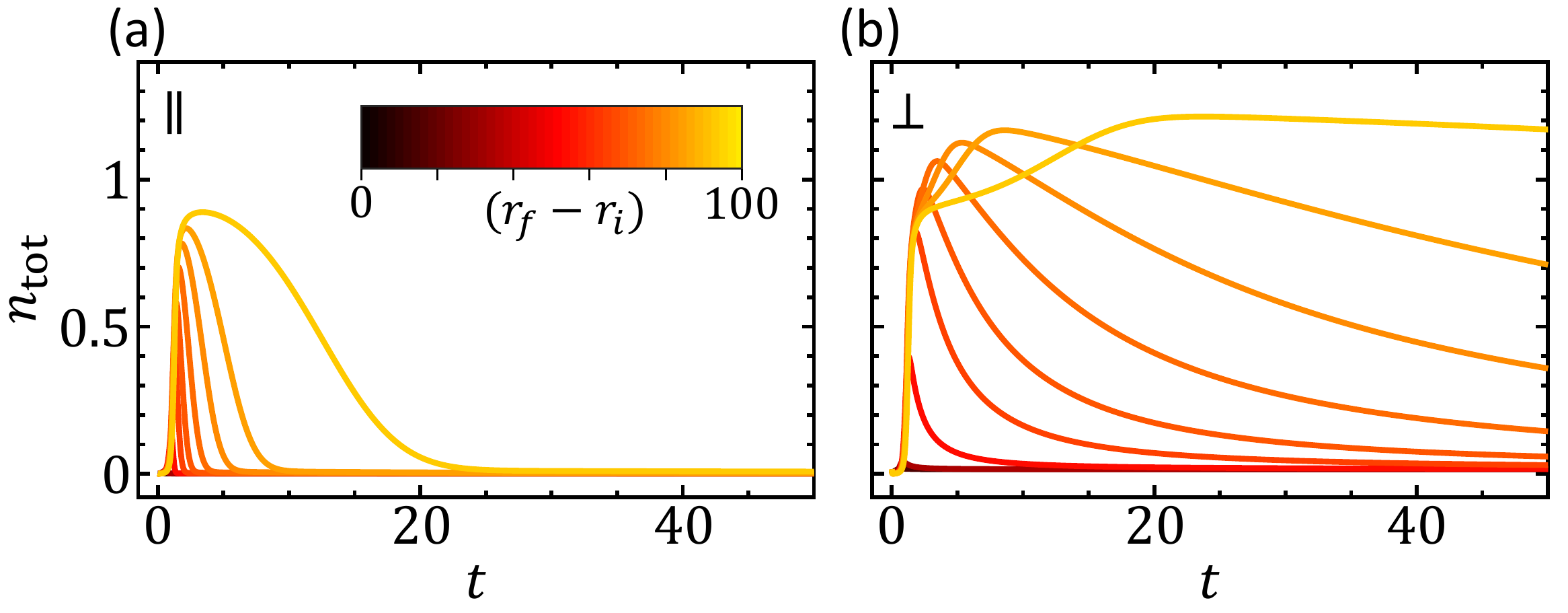}
    \caption{Separation of timescales. (a)~Long-time dynamics of the total population of longitudinal modes, $n^{\parallel}_{\rm tot}(t)$. (b)~The same for transverse modes, $n^{\perp}_{\rm tot}(t)$. When $n^{\parallel}_{\rm tot}$ is nearly fully recovered, $n^{\perp}_{\rm tot}$ approaches its maximum.}
    \label{Fig4}
\end{figure}

Because of the additional correction to the quadratic term for the longitudinal correlations in Eq.~\eqref{eqn::dt_Dpar}, the subsequent evolution -- \emph{mode decoupling} -- is very different for the longitudinal and transverse modes; see Fig.~\ref{Fig4}. The longitudinal correlations start to relax back to the thermal equilibrium value in Eq.~\eqref{eqn::D_eq}, while the transverse modes continue to proliferate, resulting in the exponent $\alpha$, cf. Eq.~\eqref{eqn::exponents}, being positive during the third dynamical stage. Moreover, by the time when $n^{\parallel}_{\rm tot}$ is sufficiently recovered, $n^{\perp}_{\rm tot}$ is about to reach its maximum. Strong experimental evidence of this separation of timescales was reported in Refs.~\onlinecite{zong2019evidence,laulhe2017ultrafast,vogelgesang2018phase}.

Just after the mode decoupling, $n^{\perp}_{\rm tot}$ starts to slowly decrease, cf. Eq.~\eqref{eqn::delta_n_tot}, suggesting that the system enters the final relaxation stage. Note that even though lowest-momenta modes $D^{\perp}_{\bf k}$ continue to proliferate at very long times, their relative contribution to $n_{\rm tot}^{\perp}$ is suppressed by the reduced phase space of these modes, which is proportional to $k^2$. The underlying dynamics is reminiscent of an inverse particle cascade in the theory of turbulence~\cite{orioli2015universal,zakharov2012kolmogorov,nazarenko2011wave}. The main difference is that in our system the dynamics is overdamped.

In the next section, we focus on the last two dynamical stages, where we find that the system exhibits self-similar scalings in time and space, manifesting power-law-like, as opposed to exponential, behavior.

\section{Universality in the intrinsic dynamics}
\label{sec:Universality}

\begin{figure}
    \centering
    \includegraphics[width=1\linewidth]{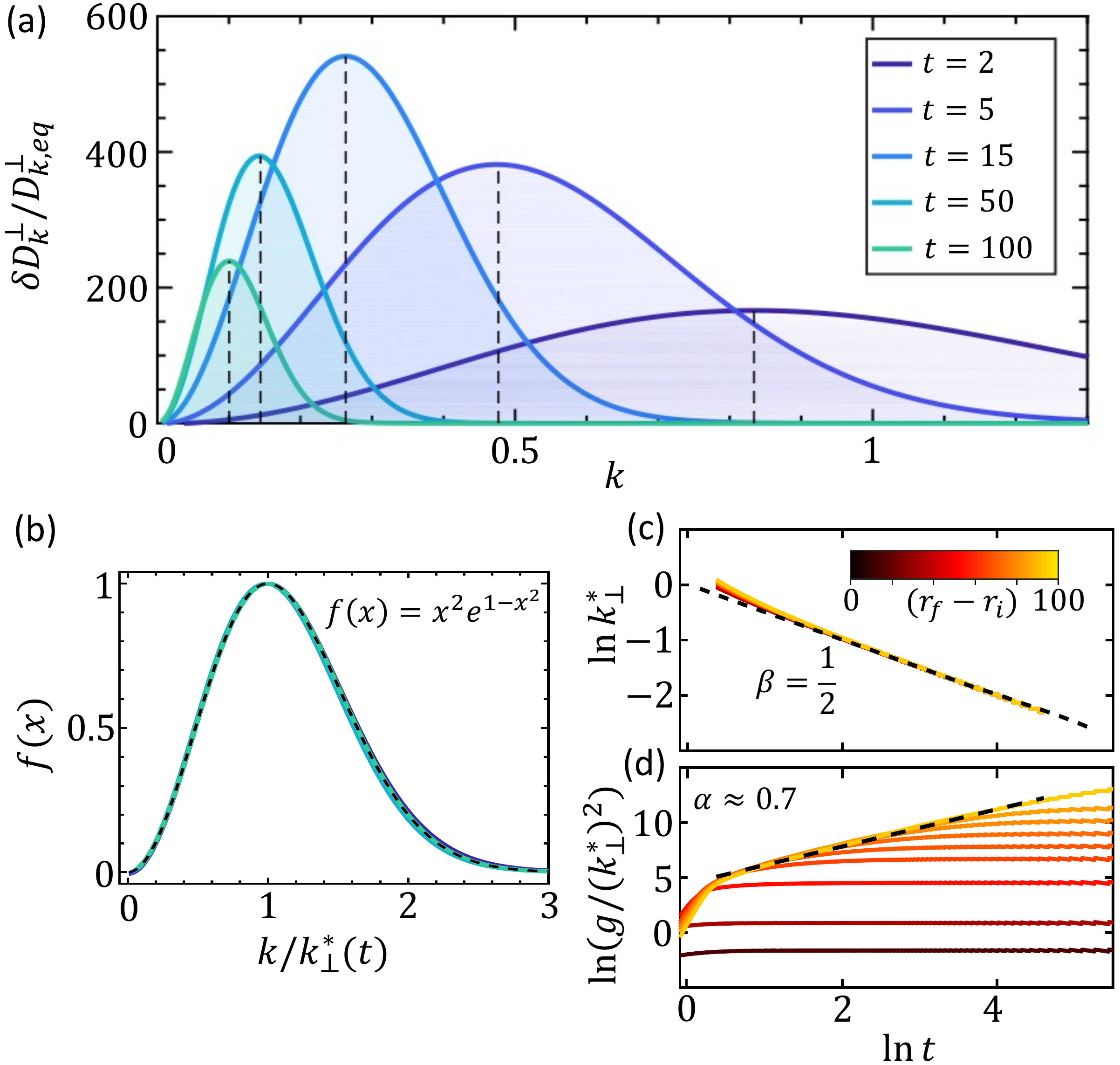}
    \caption{Long-time self-similarity. (a)~Time dependence of the change in transverse momentum distribution $\delta D^{\perp}_{\bf k}$ normalized by the equilibrium value in Eq.~\eqref{eqn::D_eq}. Quench strength is set to be $(r_f-r_i) = 80.$ Dashed lines track the position of the peak, $k_{\perp}^*(t)$; $g(t)$ corresponds to the peak height. (b)~Rescaled curves collapse into $f(x)$ [Eq.~\eqref{eqn::scaling_D_perp}]. (c)~Evolution of $k_\perp^*(t)\sim t^{-\frac{1}{2}}$ at different quench strengths. Note that $k_\perp^*(t)$ does not depend on quench. (d)~The same for the scaling function $g(t)$. From this figure we extract $\alpha \approx 0.7$ in the third dynamical stage and $\alpha = -1$ in the final stage [Eq.~\eqref{eqn::exponents}].
\label{Fig2}}
\end{figure}

Our discovery of self-similarity can be summarized in the following equations. The distribution function of the Goldstone modes follows 
\begin{eqnarray}
   \delta D_{\bf k}^{\perp}(t) \simeq \frac{g(t)}{k^2}  f(k/{k^*_\perp(t)}),\label{eqn::scaling_D_perp}
\end{eqnarray}
where $\delta D_{\bf k}^{\perp}(t) \equiv (D_{\bf k}^{\perp}(t) - D_{{\bf k}, eq}^{\perp})$ and $D_{{\bf k}, eq}^{\perp}$ is the pre-pulse equilibrium distribution given by Eq.~\eqref{eqn::D_eq}. The form in Eq.~\eqref{eqn::scaling_D_perp} is similar to the one in Eq.~\eqref{eqn:first}, though written in momentum space; $\xi_{\perp}(t) \equiv ({k^*_\perp(t)})^{-1}$ represents the emergent time-dependent length scale. We also identify the scaling relations
\begin{eqnarray}
    g(t)\sim t^{\alpha},\, k^*_\perp \sim t^{-\beta}.\label{eqn::exponents}
\end{eqnarray}
Both power-law exponents $\alpha,\, \beta$ and the function $f(x)$ are universal. We find that $\beta = \frac{1}{2}$; $\alpha \approx 0.7$ at early times while $\alpha = -1$ in the final relaxation stage. The scaling functions $f(x)$, $k^*_\perp(t)$, and $g(t)$ are shown in Fig.~\ref{Fig2}. In a model where one retains dissipation but neglects the noise coming from the bath, an exponential behavior is expected instead~\cite{mondello1992scaling}.

We first explore the implications of the self-similarity in Eq.~\eqref{eqn::scaling_D_perp} on the experimental phenomenology. Prior to the arrival of the pump pulse, the system possesses long-range coherence manifested in the macroscopic homogeneous OP $\phi$ and divergent transverse correlation length $\xi_\perp = \infty$. The laser pulse depletes this coherence. Eq.~\eqref{eqn::scaling_D_perp} suggests that as the system evolves towards equilibrium, it develops a finite correlation length $\xi_\perp(t)$ that slowly grows in a diffusive manner~\cite{bray2002theory} $\xi_\perp(t) \sim \sqrt{t}$, consistent with recent experiments~\cite{laulhe2017ultrafast,vogelgesang2018phase}. This physical picture explains the broadening of diffraction peaks observed long after the arrival of the pulse. The slowing-down of the OP recovery can also be deduced from Eq.~\eqref{eqn::scaling_D_perp}. The system enters the final dynamical stage with $g(t) \simeq A_Q t^{-1}$, where $A_Q$ is a constant of proportionality that monotonically increases with the quench strength. By contrast, as shown in Fig.~\ref{Fig2}c, $k^*_\perp (t)$ does not depend on the quench. Therefore, the cumulative effect, expressed in the change of the population of transverse modes $\delta n_{\rm tot}^\perp$, behaves as
\begin{equation}
    \delta n_{\rm tot}^\perp \equiv \int \frac{d^3 {\bf k}}{(2\pi)^3} \delta D_{\bf k}^{\perp}(t) \sim A_Q t^{-3/2}, \label{eqn::delta_n_tot}
\end{equation}
i.e. as a {\it power-law}. Since the transverse modes dominate the long-time dynamics, from Eq.~\eqref{eqn::delta_n_tot} it follows that characteristic recovery time $\tau_{\rm rec}\sim A_Q^{2/3}$ is a monotonically increasing function of the quench strength (Fig.~\ref{Fig3}b).

In the following, we offer a formal derivation of all the long-time power-law exponents: $\beta = \frac{1}{2}$, $\alpha = -1$, $d_\phi = \frac{3}{2}$, and $d_{r} = \frac{5}{2}.$ Re-establishing the long-range coherence, which is depleted by the laser pulse, is the {\it slowest} process that happens in the system, $k^*_\perp \sim t^{-\beta}$. Motivated by the numerical results in Figs.~\ref{Fig3} and~\ref{Fig2}, let us assume that $d_r > 2\beta$, which will turn out to be consistent with the subsequent derivation. We note that the most relevant transverse modes are the ones with wave vectors close to $k^*_\perp$, cf. Fig.~\ref{Fig2}. For these modes, we can safely neglect fast $r_{\rm eff}\sim t^{-d_r}$ in Eq.~\eqref{eqn::dt_Dperp} compared to slow $(k^*_\perp)^2\sim t^{-2\beta}$, resulting in a simple diffusion-like equation with the solution $\delta D^\perp_k = A_k \exp(-2\Gamma k^2 t),$ where $A_k$ is yet an unknown function of $k$. As supported by Fig.~\ref{Fig3}d, $\delta D^\perp_{k}(t)$ does not diverge for $k\to 0$. One may then Taylor-expand $A_k$ as $A_{k} = A_0 + A_2 k^2 + A_4 k^4 + \dots$ The relevant $k$ vectors, the ones in the vicinity of $k^*_\perp(t)$, are small at long times, and, thus, it is safe to leave only the dominant harmonic $A_0$ in this expansion, i.e. $\delta D^\perp_k \sim \exp(-2\Gamma k^2 t),$ consistent with $\beta = \frac{1}{2}$ and $\alpha = -1$, cf. also Fig.~\ref{Fig2}b.

To extract the value of $d_r$, we need to consider the interplay between the OP and transverse correlations (longitudinal correlations are discussed in Appendix~\ref{appendix_B}). Assuming that at long times $r_{\rm eff}\sim t^{-d_r}$, the equation of motion~(\ref{eqn::dt_phi}) reads $\frac{d}{dt} \delta \phi = -\Gamma \, r_{\rm eff}(t) \, \phi \sim t^{-d_r},$ where we implied that $\phi(t) = \phi_{0} + \delta \phi(t)$ is already close to its equilibrium value $\phi_{0}$. By integrating the above equation, we obtain $\phi^2(t) \approx \phi_{0}^2 + C t^{-d_\phi}$, where $C$ is some constant and $d_{\phi} = d_r - 1$. Note that since $\phi^2(t)$ enters the definition of $r_{\rm eff}(t)$, cf. Eq.~(\ref{eqn::r_eff}), the more dominant scaling $t^{-d_r +1}$ from the OP must be compensated by the transverse correlations. On the other hand, from the previous paragraph, we deduce $\delta n_{\rm tot}^\perp\sim t^{\alpha - \beta}$, and, therefore:
\begin{eqnarray}
   \alpha - \beta = -d_r +1 \Rightarrow d_r = 1 + \beta - \alpha = \frac{5}{2}.
\end{eqnarray}
This result also gives $d_\phi = d_r - 1 = \frac{3}{2}$. The above analysis has explained all long-time scalings. Note, however, that the self-similarity in Eq.~\eqref{eqn::scaling_D_perp} settles much earlier than the final relaxation stage. It is striking that the functional form of $f(x)$, cf. Eq.~\eqref{eqn::scaling_D_perp}, is the same for the last two dynamical stages (Fig.~\ref{Fig2}), an interesting feature that warrants further investigation.

It is worth mentioning that there are two well-known scenarios -- (i)~critical ageing dynamics~\cite{janssen1989new,calabrese2005ageing,henkel2011non,tauber2014critical} and (ii)~phase ordering kinetics~\cite{bray2002theory,mazenko1985instability,zannetti1993transverse} -- where self-similar dynamics is well established in the overdamped $O(N)$ model. Scenario (i) focuses on a temperature quench specifically to a critical point, which is not applicable here. By contrast, scenario (ii) shares some similarities with the above physical picture. The usual phase ordering kinetics is concerned with a quench from the disordered phase to a temperature below $T_c$, which parallels our quenching protocol, where the OP becomes initially suppressed and then slowly recovers to its equilibrium value. Another resemblance is that the two scenarios share the same exponent $\beta = \frac{1}{2}$. However, in the phase ordering kinetics, the classical field expectation value cannot develop, as trivially follows from Eq.~\eqref{eqn::dt_phi}, and the origin of the self-similarity is due to the growth of large phase-coherent competing domains. By contrast, the self-similarity in our case is due to the interplay between the nonzero OP $\phi$ and proliferating fluctuations, resulting in new exponents $d_\phi = \frac{3}{2}$ and $d_r = \frac{5}{2}$. The initial conditions also seem to be very different: In the phase-ordering case, one often starts from high temperatures with profound thermal fluctuations, cf. Eq.~\eqref{eqn::D_eq}; here, at the time when the OP becomes suppressed, fluctuations become depleted instead of proliferated.

\section{Discussion and Outlook}
\label{sec:discussion}

This section is devoted to addressing the applicability of our results to real experiments. In Section~\ref{sec:intro}, we motivated our study with recurring phenomenology observed in a wide variety of materials. We believe that our framework describes the underlying physics of these systems qualitatively; however, due to several features specific to each of these materials, it may alter the quantitative description. Here are some sources for potential quantitative disagreement. First, our theory is based on (3+1)-dimension, but most of those materials are either quasi-one-dimensional, such as K$_{0.3}$MoO$_3$~\cite{tomeljak2009dynamics, Huber2014}, or quasi-two-dimensional, including rare-earth tritellurides~\cite{schmitt2011ultrafast,zong2019evidence,kogar2019light,zong2019dynamical,ruan2019}, TiSe$_2$~\cite{MohrVorobeva2011}, and cuprates~\cite{mitrano2019}. Second, in multiple cases, such as Ni$_2$MnGa~\cite{mariager2014structural}, several distinct, potentially competing orders are essential. Third, systems with additional (often approximate) symmetries, such as magnetic Sr$_2$IrO$_4$~\cite{dean2016ultrafast}, are beyond the non-conserved OP evolution studied here. Fourth, some experiments reported coherent Higgs-like oscillations~\cite{zong2019evidence,tomeljak2009dynamics, Huber2014, mariager2014structural}, which are neglected in the overdamped dynamics considered here. A related point is that the model-A formalism disregards the so-called short-time dynamical slowing-down~\cite{zong2019dynamical}. Given these complications, it seems quite striking that these materials demonstrate similar behavior in pump-probe experiments.

Our work suggests that behind this common phenomenology is a simple, intuitive interpretation where slow overpopulated Goldstone modes dominate the system evolution after photoexcitation. We make a few observations to address the complications noted in the previous paragraph. First, in quasi-1D or 2D systems, fluctuations are expected to be more significant compared to 3D. Hence, we anticipate a similar, if not more pronounced proliferation of Goldstone modes. The strong spatial anisotropy can lead, though, to a different scaling exponent, as was recently observed in Ref.~\onlinecite{mitrano2019}. Second, one can straightforwardly extend the present framework to account for several orders~\cite{sun2019transient,schaefer2014collective}, and it would be intriguing to see how enhanced fluctuations enter the interplay between different orders. Third, in ferromagnetic systems, conservation of the OP should lead to additional slowing-down of the long-wavelength excitations. We leave the discussion of the conserved OP dynamics for future work. Fourth, coherent dynamics~\cite{gagel2014universal,gagel2015universal}, such as Higgs-like oscillations, is not expected to dictate the evolution for a strong pulse because these oscillations are expected to be overdamped. From a practical point of view, the model-A dynamics should also be a good approximation to describe experiments that have an insufficient time resolution to detect oscillatory behavior. On experimental side, it is essential to verify our interpretation.

Finally, we note that a variety of time-resolved experiments could be performed to test our predictions. Examples include electron or x-ray diffuse scattering \cite{chase2016,wall2018,stern2018}, resonant inelastic x-ray scattering \cite{mitrano2019}, and Brillouin scattering \cite{demokritov2001}. These experiments give access to momentum- and/or energy-resolved dynamics of bosonic excitations related to OP, so one may specifically search for signatures of: (i)~non-thermal population of the transverse modes, (ii)~the self-similarity encoded in Eq.~\eqref{eqn::scaling_D_perp}, and (iii)~different dynamical stages after photoexcitation (Fig.~\ref{Fig1}c).

\begin{acknowledgments}
The authors would like to thank A. Kogar,  B.V. Fine, A.E. Tarkhov, A. Bedroya, V. Kasper, S.L. Johnson, J. Rodriguez-Nieva, J. Marino, A. Schuckert, A. Cavalleri, G. Falkovich, and Z.-X. Shen for fruitful discussions. N.G. and A.Z. acknowledge support from the U.S. Department of Energy, BES DMSE and the Skoltech NGP Program (Skoltech-MIT  joint  project). P.E.D., M.H.M., and E.D. were supported by the Harvard-MIT Center of Ultracold Atoms, AFOSR-MURI Photonic Quantum Matter (award FA95501610323), and DARPA DRINQS program (award D18AC00014).
\end{acknowledgments}

\bibliography{cdw_lib}

\begin{thebibliography}{68}
\bibitem [{Rem()}]{Remark1}%
  \BibitemOpen
  \href@noop {} {}\bibinfo {note} {Since OP degrees of freedom are composed of
  electrons and phonons, the introduction of three subsystems might be
  confusing. For this reason, in CDW systems one sometimes splits the lattice
  into two groups: (i) phonons associated with the OP and (ii) the rest of the
  lattice. One then describes a non-equilibrium state by assigning different
  temperatures to each subsystem.}\BibitemShut {Stop}%
\end{thebibliography}

\appendix

\section{Derivation of the equations of motion}
\label{appendix:A}
Here we provide details of the derivation of the main Eqs.~(\ref{eqn::dt_phi})--(\ref{eqn::r_eff}). 

\textbf{Dynamics of $\phi$.} Evolution of the field $\phi = \frac{1}{\sqrt{V}}\phi_{1,{\bf q} = 0}$ can be obtained from:
\begin{equation}
\partial_t \mean{\phi_{\alpha, {\bf q}}}_t = \int D[\phi] \phi_{\alpha, {\bf q}} \partial_t {\cal P}([\phi],t) = -\Gamma \mean{\frac{\delta {\cal F}}{\delta \phi_{\alpha,-{\bf q}} }},\label{eqn::S2}
\end{equation}
where in the last equality we used the Fokker-Planck Eq.~(\ref{eqn::F_P}), and integration by parts. The latter derivative can be calculated from Eq.~(\ref{eqn::F}):
\begin{equation}
    \frac{\delta {\cal F}}{\delta \phi_{\alpha,-{\bf q}}} = (r + K{\bf q}^2)\phi_{\alpha, {\bf q}} + \frac{4u}{V} \sum_{{\bf k_1},{\bf k_2}} \phi_{\beta,{\bf k_1}}\phi_{\beta,{\bf k_2}}\phi_{\alpha,{\bf q- k_1-k_2}}.
\end{equation}
Using Wick's theorem and leaving only terms up to the leading order in $1/N$, we obtain
\begin{eqnarray}
   \mean{ \sum_{{\bf k_1},{\bf k_2}} \phi_{\beta,{\bf k_1}}\phi_{\beta,{\bf k_2}}\phi_{1,{\bf - k_1-k_2}}} \approx \phi_{1,{\bf q} = 0}^3 + \notag{}\\
   +\phi_{1,{\bf q} = 0} \sum_{\bf k} (D^{\parallel}_{\bf k} + (N-1)D^{\perp}_{\bf k}). \label{eqn::S4}
\end{eqnarray}
Combining Eq.~(\ref{eqn::S2}) and Eq.~(\ref{eqn::S4}) we arrive at Eq.~(\ref{eqn::dt_phi}) of the main text.

\textbf{Dynamics of the correlators.} Applying the same trick as above, we derive:
\begin{eqnarray}
    &&\partial_t \mean{\phi_{\alpha,{\bf k}}\phi_{\alpha,-{\bf k}}}_c = 2 T \Gamma - 2 \Gamma \times \notag{}\\
    &&\times \left[ \mean{\phi_{\alpha,{\bf k}} \frac{\delta F}{\delta \phi_{\alpha,{\bf k}}}} - \mean{\phi_{\alpha,{\bf k}}} \mean{\frac{\delta F}{\delta \phi_{\alpha,{\bf k} }}} \right].\label{eqn::S5}
\end{eqnarray}
For the case of the transverse component, in the leading in $1/N$ order we obtain:
\begin{eqnarray}
   && \mean{\phi_{\alpha, {\bf k}}\sum_{{\bf k_1},{\bf k_2}} \phi_{\beta,{\bf k_1}}\phi_{\beta,{\bf k_2}}\phi_{\alpha,{\bf -k- k_1-k_2}} } \notag{} \approx\\
    && \approx D^\perp_{\bf k} \left( \phi_{1,{\bf q}=0}^2 + \sum_{\bf q} (D^{\parallel}_{\bf q} + (N-1)D^{\perp}_{\bf q}) \right). \label{eqn::S6}
\end{eqnarray}
Combining Eq.~(\ref{eqn::S5}) and Eq.~(\ref{eqn::S6}) we arrive at Eq.~(\ref{eqn::dt_Dperp}) of the main text. For the case of the longitudinal component, similarly to the above discussion we get
\begin{eqnarray}
   &&\sum_{{\bf k_1},{\bf k_2}}\Big( \mean{\phi_{1,{\bf k}} \phi_{\beta,{\bf k_1}}\phi_{\beta,{\bf k_2}}\phi_{1,{\bf -k- k_1-k_2}} } - \notag{}\\
   &&-\mean{\phi_{1,{\bf k}}}\mean{\phi_{\beta,{\bf k_1}}\phi_{\beta,{\bf k_2}}\phi_{1,{\bf -k- k_1-k_2}} } \Big) \notag{} \approx\\
   && \approx D^\parallel_{\bf k} \left( 3\phi_{1,{\bf q}=0}^2 + \sum_{\bf q} (D^{\parallel}_{\bf q} + (N-1)D^{\perp}_{\bf q}) \right).
\end{eqnarray}
This equation leads to Eq.~(\ref{eqn::dt_Dpar}).

\section{Longitudinal correlations at long times}
\label{appendix_B}

\begin{figure}[hbp!]
    \centering
    \includegraphics[width=1\linewidth]{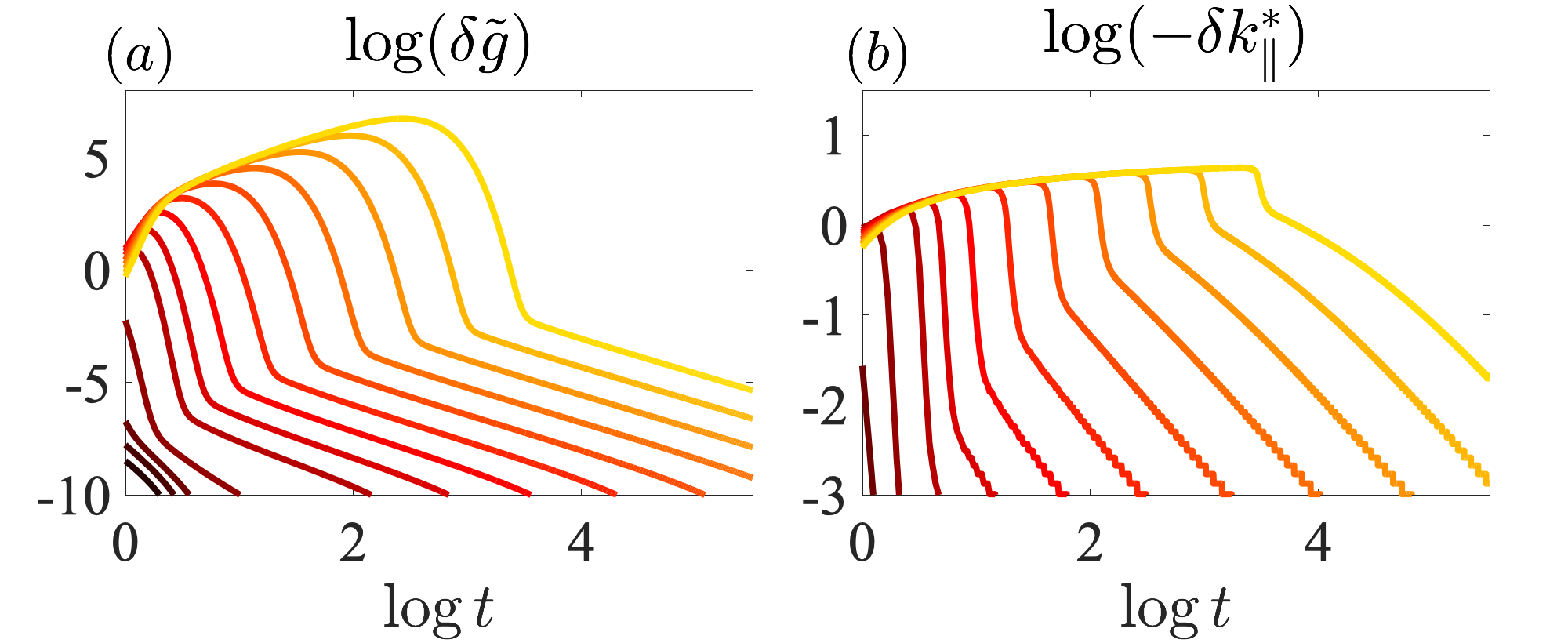}
    \caption{(a) evolution of the scaling function $\delta\tilde{g}(t)\equiv \tilde{g}(t) - \tilde{g}_{\rm eq}$ for different quenches. (b) the same for the longitudinal wave vector $\delta k^*_\parallel(t)\equiv k^*_\parallel(t) - k^*_{\parallel, {\rm eq}}$. The second (inflation) and the third (mode decoupling) stages of the overall dynamics are clearly seen. At long times, both functions scale as $\delta\tilde{g}(t),\, \delta k^*_\parallel(t) \sim t^{-d_{\phi}}$.}
    \label{fig:univ_longitudinal}
\end{figure}

During the evolution, the longitudinal correlation function $D^{\parallel}_{{\bf k}}$ remains bell-shaped with a maximum at $k = 0$ suggesting to define $\tilde{g}(t) = D^{\parallel}_{k = 0}(t)$ and $k^*_\parallel(t)$ to be the wave vector corresponding to half width at half maximum in $D^{\parallel}_{{\bf k}}$. Notably, both functions at long times behave as $\tilde{g}(t),\, k^*_\parallel(t) \sim t^{-d_{\phi}}$ -- see Fig.~\ref{fig:univ_longitudinal}. We also observe that this power-law exponent implies that the longitudinal correlations exhibit the leading scaling, i.e. these modes should not be entirely ignored. 

To explain the above observation, we note that at long times, when the order parameter $\phi(t) = \phi_0 + \delta \phi$ is already close to be recovered, the equation of motion~\eqref{eqn::dt_Dpar} can be approximated to (we fix $K = 1$ for convenience)
\begin{equation}
    \frac{d \delta D^{\parallel}_{{\bf k}}}{dt} \approx -32 \Gamma u \phi_0 \delta \phi D^{\parallel}_{{\bf k}, {\rm eq}} - 2\Gamma ( k^2 + 8u\phi_0^2) \delta D^{\parallel}_{{\bf k}},
\end{equation}
where $D^{\parallel}_{{\bf k}}(t) = D^{\parallel}_{{\bf k}, {\rm eq}} + \delta D^{\parallel}_{{\bf k}}(t)$ and we disregarded fast $r_{\rm eff}(t) \sim t^{-d_r}$ compared to slow $\delta \phi(t)\sim t^{-d_{\phi}}$ ($0 <d_{\phi}< d_{r}$). The above equation can be solved analytically. Indeed, substituting $\delta D^{\parallel}_{{\bf k}}(t) = {\rm e}^{ - 2\Gamma (k^2 + 8u\phi_0^2) t} h_{{\bf k}}(t),$ we obtain the following equation on $h_{{\bf k}}(t)$:
\begin{equation}
    \frac{d h_{{\bf k}}}{dt} = -32\Gamma u\phi_0 \delta \phi D^{\parallel}_{{\bf k}, {\rm eq}} {\rm e}^{ 2\Gamma (k^2 + 8u\phi_0^2) t}.
\end{equation}
Integration of this equation gives
\begin{equation}
    h_{{\bf k}}(t) = h_{{\bf k}}(t_0) + \frac{C}{k^2 + 8u\phi_0^2} \int\limits_{t_0}^t dt'  \frac{{\rm e}^{ 2\Gamma (k^2 + 8u\phi_0^2) t'}}{(t')^{d_{\phi}}},
\end{equation}
where $C$ is some constant. We, therefore, conclude that
\begin{equation}
    \delta D^\parallel_{\bf k} = \delta D^{\parallel, (1)}_{\bf k} + \delta D^{\parallel, (2)}_{\bf k},
\end{equation}
where $\delta D^{\parallel, (1)}_{\bf k}(t) =h_{{\bf k}}(t_0) {\rm e}^{ - 2\Gamma (k^2 + 8u\phi_0^2) t}$ decays exponentially in time, whereas
\begin{equation}
    \delta D^{\parallel, (2)}_{\bf k} \sim  \frac{{\rm e}^{ - 2\Gamma (k^2 + 8u\phi_0^2) t}}{k^2 + 8u\phi_0^2} \int\limits_{t_0}^t dt'  \frac{{\rm e}^{ 2\Gamma (k^2 + 8u\phi_0^2) t'}}{(t')^{d_{\phi}}}\label{eqn::D_par_2}
\end{equation}
is potentially important. 

At long times $t\rightarrow \infty$, we observe that
\begin{equation}
    F(t) \equiv \int\limits_{t_0}^t dt'  \frac{{\rm e}^{ a t'}}{(t')^{b}} \sim \frac{{\rm e}^{ a t}}{t^{b}},\, a,b>0.\label{eqn::general_long_time}
\end{equation}
Indeed, by differentiating $F(t)$ we note that it satisfies
\begin{equation}
    \frac{d F}{dt} = \frac{{\rm e}^{ a t}}{t^{b}}.
\end{equation}
By substituting $F(t) = {\rm e}^{ a t} p(t)$ we separate rapid exponential growth from slow power-law-like dynamics encoded in $p(t)$:
\begin{equation}
    \frac{d p}{d t} + a p = \frac{1}{t^{b}}.
\end{equation}
From this equation, we finally see that $p \sim t^{-b}$ (as long as $a \neq 0$). Combining Eqs.~(\ref{eqn::D_par_2}) and~(\ref{eqn::general_long_time}), we conclude that 
\begin{equation}
    \delta D^{\parallel, (2)}_{\bf k} \sim  \frac{ t^{-d_\phi}}{k^2 + 8u\phi_0^2},
\end{equation}
i.e. indeed $\delta D^{\parallel}_{\bf k}$ gets power-law-like contribution with the leading exponent. For completeness, we also note that
\begin{equation}
    \delta n^{\parallel, (2)}_{\rm tot} = \int\frac{d^3 {\bf k}}{(2\pi)^3} \delta D^{\parallel, (2)}_{\bf k} \sim t^{-d_\phi}
\end{equation}
also exhibits the same scaling.

\end{document}